\documentclass[amsmath,amssymb,twocolumn,prd]{revtex4}
\usepackage{pstricks}
\usepackage{graphicx}

\def\al{\alpha}
\def\be{\beta}
\def\ga{\gamma}
\def\de{\delta}
\def\ep{\epsilon}

\def\ze{\zeta}
\def\et{\eta}
\def\th{\theta}

\def\ka{\kappa}
\def\la{\lambda}

\def\rh{\rho}

\def\ph{\phi}

\def\ch{\chi}

\def\om{\omega}

\def\De{\Delta}

\def\vev#1{\langle {#1}\rangle}

\def\half{{\tfrac12}}
\def\lsim{\mathrel{\rlap{\lower4pt\hbox{\hskip1pt$\sim$}}
    \raise1pt\hbox{$<$}}}
\def\gsim{\mathrel{\rlap{\lower4pt\hbox{\hskip1pt$\sim$}}
    \raise1pt\hbox{$>$}}}
\def\sqr#1#2{{\vcenter{\vbox{\hrule height.#2pt
         \hbox{\vrule width.#2pt height#1pt \kern#1pt
         \vrule width.#2pt}
         \hrule height.#2pt}}}}
\newcommand{\beq}{\begin{equation}}
\newcommand{\eeq}{\end{equation}}
\newcommand{\bea}{\begin{eqnarray}}
\newcommand{\eea}{\end{eqnarray}}
\newcommand{\rf}[1]{(\ref{#1})}
\def\etal{{\it et al.}}

\def\re{{\rm Re}}
\def\im{{\rm Im}}

\def\vec#1{\boldsymbol #1}
\def\unit#1{\hat{\boldsymbol #1}}

\def\db{\displaybreak[0]}

\def\voc{\mathrel{\rlap{\lower0pt\hbox{\hskip1pt{$c$}}}\raise3pt\hbox{$\neg$}}}
\def\sk#1#2#3{#1^{(#2)}_{#3}}
\def\cjm#1#2#3{c^{(#1)}_{(#2)#3}}
\def\cF{\sk{(c_F^{(d)})}{0E}{njm}}
\def\cFdnjm#1#2{\sk{(c_F^{(#1)})}{0E}{#2}}
\def\cI{\cjm{d}{I}{jm}}
\def\cIdjm#1#2{\cjm{#1}{I}{#2}}
\def\cf{\sk{(\voc_F^{(d)})}{0E}{njm}}
\def\cfdnjm#1#2{\sk{(\voc_F^{(#1)})}{0E}{#2}}

\def\Mjm#1#2#3{{\cal M}^{(#1)}_{(#2)#3}}
\def\Mc{\Mjm{d}{c_F}{njm}}
\def\Mct{\Mjm{d}{\voc_F}{njm}}
\def\MI{\Mjm{d}{I}{jm}}
\def\Mcdnjm#1#2{\Mjm{#1}{c_F}{#2}}

\def\MIdjm#1#2{\Mjm{#1}{I}{#2}}
\def\Mccav{{\cal M}^{(d)\, \rm cav}_{(c_F)njm}}

\def\Mccavdnjm#1#2{{\cal M}^{(#1)\, \rm cav}_{(c_F)#2}}

\def\Mclab{{\cal M}^{(d)\, \rm lab}_{(c_F)njm}}

\def\Mclabdnjm#1#2{{\cal M}^{(#1)\, \rm lab}_{(c_F)#2}}

\def\Mcarm{{\cal M}^{(d)\, \rm arm}_{(c_F)njm}}

\def\McarmA{{\cal M}^{(d)\, \rm A}_{(c_F)njm}}

\def\McarmB{{\cal M}^{(d)\, \rm B}_{(c_F)njm}}

\def\McarmC{{\cal M}^{(d)\, \rm C}_{(c_F)njm}}

\def\McarmD{{\cal M}^{(d)\, \rm D}_{(c_F)njm}}

\def\Mzero#1#2{{\mathcal U}^{(#1)}_{#2}}
\def\Mtwo#1#2{{\mathcal V}^{(#1)}_{#2}}
\def\MzeroA#1#2#3{{\mathcal U}^{(#2)#1}_{#3}}
\def\MtwoA#1#2#3{{\mathcal V}^{(#2)#1}_{#3}}

\def\sE{\underline E}
\def\sD{\underline D}

\def\sH{\underline H}

\def\ss{\underline s}
\def\calE{\mathcal E}
\def\syjm#1#2{\phantom{}_{#1}Y_{#2}}

\begin{document}

\title{Optical-cavity tests of higher-order Lorentz violation}

\author{Matthew Mewes}
\affiliation{
  Department of Physics and Astronomy, Swarthmore College,
  Swarthmore, Pennsylvania 19081, USA}

\begin{abstract}
  The effects of
  Lorentz-violating operators
  of nonrenormalizable dimension
  in optical resonate cavities
  are studied.
  Optical-frequency experiments are shown
  to provide sensitivity to 
  nondispersive nonbirefringent
  violations that is many orders
  of magnitude beyond current constraints
  from microwave cavities.
  Existing experiments based on
  Fabry-P\'erot and ring
  resonators are considered
  as illustrations.
\end{abstract}

\maketitle

\section{introduction}

Lorentz invariance is a
cornerstone of modern physics.
However, the possibility that Planck-scale
physics may give rise to minute defects
in this fundamental principle \cite{ks}
has motivated numerous experimental
tests of Lorentz symmetry.
Searches for Lorentz violation
have been performed
in many different systems,
including those involving photons \cite{datatables}.
Among these are modern versions of
the classic Michelson-Morley experiment \cite{mm}.
The contemporary experiments are
based on electromagnetic resonant
cavities and provide extreme
sensitivity to potential Lorentz violation
\cite{microwave,optical,cavities,schiller,peters,parker,ring}.

A theoretical framework known
as the Standard-Model Extension (SME)
provides a general field-theoretic description
of Lorentz and CPT violation at
attainable energies \cite{ck}.
The SME aids in the identification of 
experimental signatures
and in the comparison
of different measurements.
Many tests of Lorentz 
and CPT invariance have focused on the
minimal Standard-Model Extension (mSME),
which restricts attention to
Lorentz-violating operators of
renormalizable dimension in flat spacetime.
However, recent studies have extended the SME
formalism to Lorentz violation
involving curved spacetime
\cite{gravsme,gravexp}
and nonrenormalizable operators
\cite{parker,km_apjl,km09,dispersion,nonrenorm,neutrinos}.
This work examines the effects
of higher-order nonrenormalizable
electromagnetic operators
in optical resonant-cavity experiments.
The study of higher-order terms is
motivated in part by the different
physical effects they introduce,
as well as the possibility that
the dominate Lorentz violation may involve
nonrenormalizable operators only.
A more detailed discussion of the experimental
and theoretical implications can be found in
Ref.\ \cite{km09}.

The Lorentz-violating terms
in the photon sector of the SME
can be classified according
to various properties,
such as the mass dimension $d$ of the operator.
The renormalizable operators
of the mSME have dimensions $d=3,4$,
while operators of higher dimension,
$d\geq 5$, are nonrenormalizable.
The odd-dimensional operators break CPT,
in addition to breaking Lorentz invariance.

It is also useful to split 
the set of Lorentz-violating operators
into those that affect the vacuum propagation
of light and those that do not,
at leading order.
For example,
some operators result in vacuum birefringence,
which causes polarization to change as light
propagates through empty space
\cite{km_apjl,km09,km02,bire}.
Lorentz violation can also lead to dispersion,
giving rise to energy-dependent propagation speeds
\cite{km_apjl,km09,dispersion}.
Both these effects can
be tested to very high precision
using light from astrophysical sources,
where the tiny effects of Lorentz violation
are enhanced by the cosmological distances involved,
but not all forms of Lorentz violation result
in leading-order birefringence or dispersion.
The nonbirefringent nondispersive
operators are comparatively difficult to detect,
since they have little effect on light propagating
{\em in vacuo}.
They do, however,
affect electromagnetic resonances in cavities.
Cavity experiments thus provide
a class of Lorentz-invariance tests that 
complement astrophysical tests.

The nonbirefringent nondispersive
violations are controlled by the
set of camouflage coefficients \cite{km09}
and are the main focus of this work.
Camouflage coefficients exist
for even dimensions $d\geq 4$
and are invariant under CPT.
Here, we examine the prospect
of measuring higher-dimensional ($d\geq 6$)
camouflage coefficients using optical cavities.
At present,
the only bounds on these coefficients
are from a microwave-cavity experiment,
which placed constraints on combinations
of $d=6$ and $d=8$ coefficients \cite{parker}.

The sensitivity to the various camouflage coefficients
is largely determined by the frequency and
geometry of the modes excited in a cavity experiment.
For example, many experiments utilize cavities
that are symmetric under a parity transformation.
These have been found to provide direct access
to parity-even anisotropic coefficients in the SME
through direction-dependent resonant frequencies.
In contrast,
the effects from parity-odd and isotropic coefficients
only contribute through boost violations.
As a result, they enter in conjunction
with the boost velocity $\be$ for parity-odd violations
and the velocity-squared $\be^2$ for isotropic violations.
The relevant boost speeds are normally those
from the rotational and orbital
motion of the Earth, $\be\lsim 10^{-4}$,
leading to suppressed sensitivities
to parity-odd and isotropic violations.
This suppression may be overcome by use of
parity-breaking interferometers and resonators
\cite{ring,machzehnder,exirifard_ring,petroff}.
Other techniques for testing these violations
include those involving
static fields \cite{emstats},
accelerators \cite{accel},
and \v Cerenkov radiation \cite{cerenkov},
among others \cite{other}.

There are several major advantages
to optical experiments.
First, to a good approximation,
the resonances can be
taken as plane waves.
As demonstrated below,
this yields analytic expressions
for the frequency shifts,
where microwave cavities
typically require numerical calculations.
Second, parity-violating cavities,
such as ring resonators,
are comparatively easy to construct
at optical frequencies.
Finally, in general,
the sensitivity to higher-order
coefficients grows with frequency as $\om^{d-4}$.
This implies the potential for
improved sensitivities by
approximately a factor $10^{4(d-4)}$
over microwave experiments.
We therefore expect
improvements in sensitivity of
roughly eight orders of magnitude for $d=6$
and sixteen orders of magnitude for $d=8$.

This paper is organized as follows.
In Sec.\ \ref{sec_theory},
we discuss some basic theory that
is common to all optical-cavity experiments.
The sections that follow provide analyses 
of several recent experiments \cite{schiller,peters,ring},
as illustrations.
In Sec.\ \ref{sec_fabry_perot},
we derive the sensitivity 
of crossed Fabry-P\'erot cavities
to parity-even Lorentz violations.
A parity-odd experiment based
on a ring resonator is considered in
Sec.\ \ref{sec_ring}.

\section{basic theory} \label{sec_theory}

In this section,
we establish the basic theory behind
most cavity-based experiments.
A more detailed explanation
of the nonrenormalizable terms considered here
can by found in Ref.\ \cite{km09}.
Here, we summarize the parts that
are relevant to optical experiments.

The basic idea behind a typical 
resonator experiment is to look
for a shift in resonant frequency
due to Lorentz violation \cite{km02}.
In general,
this shift will be frame dependent,
leading to variations in frequency
with changes in cavity orientation or speed.
This work focuses on changes in orientation
due to active rotations of the cavity
in the laboratory and to the rotation
of the Earth throughout the day.

We begin by focusing
on nonbirefringent Lorentz violations,
which are characterized
by coefficients $\cF$ in the SME.
These coefficients are
associated with even-dimensional
operators, $d = \text{even} \geq 4$.
The $n$ index controls the frequency/wavelength dependence
and is limited by $0\leq n \leq d-2$.
Rotational properties are
determined by the usual
angular-momentum indices $j$ and $m$,
where $j=n,n-2,\ldots,\geq 0$
and $|m|\leq j$.
The shift in resonant frequency
due to nonbirefringent terms
in the SME takes the form
\beq
\frac{\de\nu}{\nu} = \!\!\!\sum_{dnjmm'}\!\!\! \Mclabdnjm{d}{njm} 
e^{im\ph + im'\om_\oplus T_\oplus} d^{(j)}_{mm'}(-\ch)
\cFdnjm{d}{njm'} \, .
\eeq
The $\Mclab$ are experiment-dependent factors
that determine the sensitivity of the
resonator.

It is important to specify the frame
in which various quantities are defined.
By convention, we take the
coefficients for Lorentz violation $\cF$
in the Sun-centered frame
defined in Ref.\ \cite{datatables}.
This frame is inertial to a very good approximation,
so the coefficients in this frame are constant.
We could, in principle, calculate
the $\Mc$ matrices in this same frame,
but they would then vary in time.
This is because they take the orientation
of the cavity into account, 
and the orientation relative to the Sun frame
changes with time,
giving rise to a varying $\Mc$ matrix.

Alternatively, we can account for the
change in orientation by adopting a rotating
laboratory frame within which the cavity is stationary.
In this frame, the $\Mclab$ are constants.
The changes in orientation are then incorporated
through the rotation that relates the two frames.
This rotation takes the form of the 
$e^{im\ph + im'\om_\oplus T_\oplus} d^{(j)}_{mm'}(-\ch)$
factor in the frequency shift.
The $d^{(j)}_{mm'}$ are little Wigner matrices,
and $\ch$ is the colatitude of the laboratory.
By convention, the laboratory-frame $z$
axis is vertical.
The angle $\ph$ is the angle between
the laboratory-frame $x$ axis and south.
The angle $\om_\oplus T_\oplus$ 
is the right ascension of the local zenith,
where $\om_\oplus\simeq 2\pi/(\text{23 hr 56 min})$
is the Earth's sidereal rotation rate.
Many experiments place the cavities on turntables.
In this case,
we may write $\ph = \om_\text{tt} T_\text{tt}$,
where $\om_\text{tt}$ is the turntable
rotation rate.

The $\Mclab$ constants can be found 
perturbatively using
the conventional electric field $\vec E$ for
the resonate mode being studied.
To avoid divergences in the calculation
that stem from discontinuities at the boundary
of the cavity,
we must also define a smoothed field $\vec\sE$,
which matches $\vec E$ inside the resonator,
but extends smoothly to the outside region.
Switching to momentum space,
the fields contribute to $\Mclab$
through the spin-weighted Stokes combinations
\beq
  \ss^0 = (E_+)^* \sE_+ + (E_-)^* \sE_- \ ,
  \quad
  \ss_{(\pm2)} = 2(E_\pm)^* \sE_\mp \ ,
\eeq
where
$E_\pm = \vec E\cdot(\unit\th \pm i\unit\varphi)/\sqrt2$ and 
$\sE_\pm = \vec\sE\cdot(\unit\th \pm i\unit\varphi)/\sqrt2$.
Here, $\unit\th$ and $\unit\varphi$
are the usual unit vectors associated
with the polar angle $\th$ 
and azimuthal angle $\varphi$.
In momentum space,
$\th$ is the angle between
the momentum $\vec p$ and the $z$ axis,
while $\varphi$ is the angle between
the $x$-$y$ projection of $\vec p$
and the $x$ axis.

For optical resonators,
the $\Mc$ matrix elements are given by \cite{km09}
\begin{widetext}
\begin{align}
  \Mc
  &= \frac{\om^{d-4-n}}{4\vev{U}}\int d^3p\, p^{n-2} \bigg[
    \tfrac14 (\om^2-p^2)\sqrt{\tfrac{(j+2)!}{(j-2)!}}
    \big(\syjm{+2}{jm}(\unit p)\, \ss_{(+2)}(\vec p)
    + \syjm{-2}{jm}(\unit p)\, \ss_{(-2)}(\vec p)\big)\,
    \notag\\&\quad
    +\Big((n-\tfrac{j(j+1)}{2})(\om^2-p^2) - (d-2-n)(d-3-3n)p^2 - n(n-1)p^2 \Big)\,
    \syjm{0}{jm}(\unit p)\, \ss^0(\vec p)
    \bigg] \ ,
  \label{Mc}
\end{align}
\end{widetext}
where $\syjm{s}{jm}$ are
spin-weighted spherical harmonics, and
$\vev{U} = 
\tfrac14 \int d^3p\,
(\vec E^*\cdot \vec \sD + \vec B^* \cdot \vec \sH)$
is the time-averaged energy in the cavity.
In general, there are additional terms
in $\Mc$ that arise from
longitudinal polarization.
These, however, do not contribute
in optical resonators where the
fields can be approximated
as transverse-polarized plane waves.
We illustrate how to use
Eq.\ \rf{Mc} to get the laboratory-frame
$\Mclab$ constants in the following sections.

In practice,
experiments search for the frequency
shift by comparing the resonances
of two different cavities or modes.
The beat frequency between the
two modes is then analyzed for
variations at the sidereal
and turntable rotation rates.
The beats will depend on $\De\Mclab$,
the difference between the $\Mclab$
constants for the two modes.
It is convenient to write
the beat frequency as
\beq
\frac{\nu_\text{beat}}{\nu} 
= \sum_{mm'} A_{mm'} e^{im\ph + im'\om_\oplus T_\oplus} \ ,
\label{nu_beat1}
\eeq
where
\beq
A_{mm'} = \sum_{dnj} \De\Mclabdnjm{d}{njm} 
d^{(j)}_{mm'}(-\ch) \cFdnjm{d}{njm'} \ .
\label{Amm}
\eeq
The $A_{mm'}$ factors are
linear combinations
of coefficients for Lorentz violation,
which satisfy $A^*_{mm'} = A_{(-m)(-m')}$.
They represent the complex combinations of
coefficients a given experimental
configuration can access.

Two basic strategies can be used to
extract constraints on coefficients
for Lorentz violation.
The first is to search directly for variations at rates
$\om_{mm'} = m\om_\text{tt} + m'\om_\oplus$.
However, since $\om_\oplus$ is typically
much smaller than $\om_\text{tt}$,
the frequencies $\om_{mm'}$ are closely spaced around
the harmonics of the turntable rotation frequency,
making it difficult to discriminate
many of the variations in the beat frequency.

The second strategy for analysis relies
on the fact that the sidereal variations
($m'\om_\oplus$)
can be viewed as slow
modulations of the amplitudes of the harmonics
of the turntable frequency.
That is, we write \cite{parker}
\beq
\frac{\nu_{\rm beat}}{\nu} = \sum_{m\geq 0} 
\big[ C_m(T_\oplus) \cos (m\ph) + S_m(T_\oplus)\sin (m\ph) \big] \ ,
\label{nu_beat2}
\eeq
where
\begin{align}
  C_m(T) &= \!\!\sum_{m'\geq 0}\!\!
  \big[C^C_{mm'} \cos (m'\om_\oplus T) 
    + C^S_{mm'} \sin (m'\om_\oplus T) \big]\, ,
  \notag \\
  S_m(T) &= \!\!\sum_{m'\geq 0}\!\!
  \big[ S^C_{mm'} \cos (m'\om_\oplus T) 
    + S^S_{mm'} \sin (m'\om_\oplus T) \big]\, .
\label{CS}
\end{align}
The sidereal amplitudes are given by
\begin{align}
  C^C_{mm'} &= 2\, \et_m \et_{m'}\, 
  \re\big[A_{mm'} + A_{m(-m')}\big] \ ,
  \notag \\
  C^S_{mm'} &= -2\, \et_m\, 
  \im\big[A_{mm'} - A_{m(-m')}\big] \ ,
  \notag \\
  S^C_{mm'} &= -2\, \et_{m'}\, 
  \im\big[A_{mm'} + A_{m(-m')}\big] \ ,
  \notag \\
  S^S_{mm'} &= -2\, \re\big[A_{mm'} - A_{m(-m')}\big] \ ,
  \label{CSamps}
\end{align}
where $\et_0=1/2$,
and $\et_m = 1$ when $m\neq 0$.
Equation \rf{CSamps} gives real
linear combinations of coefficients for
Lorentz violation that the experiment can measure.
The theoretical analysis
is then reduced to finding
these combinations in terms of
the $\cF$ coefficients.

The above gives the sensitivity
of a cavity experiment to
the nonbirefringent 
Lorentz-violating operators
of the SME.
However, many of the 
$\cF$ coefficients lead to 
vacuum dispersion.
To get the sensitivity
to the camouflage coefficients,
we now restrict attention
to the subset of the 
$\cF$ coefficients
that are nondispersive.
These are denoted by $\cf$.
Setting all other coefficients to zero,
the camouflage coefficients
are related to $\cF$ through
\beq
\cF = \cf - \cfdnjm{d}{(n-2)jm} \ .
\eeq
They are nonzero for index values in the ranges
$d=\text{even}\geq 4$,
$0\leq n\leq d-4$,
and
$j=n,n-2,\ldots,\geq 0$.
We can find the sensitivity to the camouflage
coefficients by simply replacing
$\cF$ with $\cf$ and
$\Mc$ with
\beq
\Mct = \Mc - \Mcdnjm{d}{(n+2)jm}
\eeq
in the beat frequency.

The dimension-four operators for
Lorentz violation provide one more class
of coefficients that are nondispersive.
This special case exists because
none of the $d=4$ operators lead
to energy-dependant velocities.
Consequently, the portion of
$\cFdnjm{d}{njm}$ that is dispersive
for $d>4$ is not dispersive for $d=4$.

There exists an alternative equivalent
representation of these $d=4$ coefficients,
which naturally arises in the
analysis of plane waves.
Denoted by $\cI$,
they are related to the $\cFdnjm{4}{njm}$
coefficients through
\begin{align}
  \cFdnjm{4}{200} &= \tfrac13 \cFdnjm{4}{000} = \tfrac14 \cIdjm{4}{00} \ ,
  \notag \\
  \cFdnjm{4}{11m} &= -2 \cIdjm{4}{1m} \ ,
  \notag \\
  \cFdnjm{4}{22m} &= \cIdjm{4}{2m} \ .
\end{align}
Sensitivity to this set of coefficients
can be found by replacing $\cF$
with $\cI$ and $\Mc$ with
$\MI$ in the beat frequency.
The relation between the $\Mcdnjm{4}{njm}$
and $\MIdjm{4}{jm}$ matrix elements is
\begin{align}
\MIdjm{4}{00} &= \tfrac34 \Mcdnjm{4}{000} + \tfrac14 \Mcdnjm{4}{200} \ ,
\notag \\
\MIdjm{4}{1m} &= -2\Mcdnjm{4}{11m} \ ,
\notag \\
\MIdjm{4}{2m} &= \Mcdnjm{4}{22m} \ .
\end{align}
This completes our discussion
of the basic theory.
We consider two optical-cavity 
experiments in the following sections,
as illustrations.

\section{Fabry-P\'erot cavity} \label{sec_fabry_perot}

To find the effects of
the camouflage coefficients
for Lorentz violation in
Fabry-P\'erot cavities,
we model the field inside the cavity 
as a linearly polarized standing wave.
We start by defining a reference frame 
with its $z$ axis along the cavity axis
and the polarization in the $y$ direction.
The nonzero field components can then be taken as
$E_y(\vec x) = \calE\rh(\vec x)\sin kz$,
where we define a profile function with
$\rh = 1$ inside the beam and $\rh=0$ outside the beam.
For the smoothed field,
we simply extend the standing wave
to infinity, giving
$\sE_y(\vec x) = \calE\sin kz$
everywhere.

The next step in the calculation is to
find the $\vec p$-space versions of the fields,
$\vec E(\vec p)$ and $\vec \sE(\vec p)$.
The fact that $\sE_y(\vec x)$
is a standing plane wave
implies $\sE_y(\vec p)$ contains
contributions from $\vec p = \pm k\unit z$ only.
This makes the determination of cavity-frame
$\Mc$ matrix elements
relatively straightforward.
For example, a short calculation gives
spin-weighted Stokes parameters
\begin{align}
\ss^0(\vec p) &=  \frac{\vev{U}}{\ep} \big[\de(\vec p-k\unit z)
+\de(\vec p+k\unit z)\big]\ , \\
\ss_{(\pm2)}(\vec p) &= -\frac{\vev{U}}{\ep}
\big[\de(\vec p-k\unit z)e^{\pm 2i\varphi}
+\de(\vec p+k\unit z)e^{\mp 2i\varphi}\big] \ ,
\end{align}
in term of the energy $\vev{U}$
and the relative permittivity $\ep$
of the material filling the cavity, if present.
This leads to
\begin{widetext}
\begin{align}
  \Mccav
  &= \frac{\om^{d-4}N^{n-2}}{4\ep}
  \Big[
    \tfrac14 (N^2-1)\sqrt{\tfrac{(j+2)!}{(j-2)!}}
    \Big(\syjm{+2}{jm}(\unit z) e^{2i\varphi} + \syjm{-2}{jm}(\unit z) e^{-2i\varphi} \Big)
    \notag\\&\quad
    -\Big((n-\tfrac{j(j+1)}{2})(N^2-1) + (d-2-n)(d-3-3n)N^2 + n(n-1)N^2 \Big)\, \syjm{0}{jm}(\unit z)
    \Big]
    \notag\\&\quad
    + (\unit z \rightarrow -\unit z,\varphi \rightarrow -\varphi) \ ,
\end{align}
\end{widetext}
where we let $N=k/\om$ be the index of refraction
of the media inside the cavity.

The advantage of the cavity frame is that
the spin-weighted spherical harmonics take
a simple form for propagation
in the $\pm\unit z$ direction.
The only nonzero harmonics are
those with $m = \mp s$ for propagation
in the $\pm\unit z$ direction.
They are given by
\begin{align}
\syjm{s}{j(-s)}(\unit z) &=
\sqrt{\tfrac{2j+1}{4\pi}} (-1)^s e^{-is\varphi} \ , \\
\syjm{s}{js}(-\unit z) &=
\sqrt{\tfrac{2j+1}{4\pi}} (-1)^{s+j} e^{is\varphi} \ ,
\end{align}
provided $j\geq|s|$, as usual.
Using these identities,
we can write the cavity-frame matrices as
\beq
\Mccav = \Mzero{d}{nj} \de_{m,0} 
+ \Mtwo{d}{nj} \de_{|m|,2} \ ,
\eeq
for even values of $j$.
They are zero for odd values of $j$.
The coefficients in the above expression are 
\begin{align}
\Mzero{d}{nj} &=
-\frac{\om^{d-4}N^{n-2}}{2\ep}\sqrt{\tfrac{2j+1}{4\pi}}
\Big[(n-\tfrac{j(j+1)}{2})(N^2-1)
\notag \\ &\hspace*{-2pt}
+ (d-2-n)(d-3-3n)N^2
+ n(n-1)N^2 \Big] \ , 
\label{Mzero}
\\
\Mtwo{d}{nj} &=
\frac{\om^{d-4}N^{n-2}(N^2-1)}{8\ep}
\sqrt{\tfrac{(2j+1)(j+2)!}{4\pi(j-2)!}}\ .
\label{Mtwo}
\end{align}
These same numerical factors will appear
again in the ring-resonator case 
discussed in the next section
and generically determine the sensitivities
of optical experiments to Lorentz violation.

Notice that the above results
simplify dramatically in
an empty cavity where $N=1$.
In particular, the $m=\pm 2$ matrix
elements vanish and the only contribution
is from the $m=0$ terms.
This implies invariance under rotations
about the cavity axis,
which means the result is independent of polarization.
This stems from the fact that we are considering
effects of nonbirefringent coefficients,
which have a uniform effect on all polarizations.
In matter, when $N\neq 1$,
nonbirefringent coefficients can lead to birefringence
\cite{km09}.
Consequently,
the introduction of media can lead to
polarization dependence even when
no polarization dependence results in the vacuum.
In the present context,
this is reflected in the possibility of
azimuthal dependence from nonzero 
$\Mccav$ matrix components for $m=\pm 2$.

Before we can apply the above result,
we must transform the cavity-frame
$\Mccav$ matrix to
the standard laboratory frame,
where $z$ is vertical and 
the $x$ axis is at an angle $\ph$ from south.
In the spherical-harmonic basis,
rotations take the form
of Wigner matrices $D^{(j)}_{mm'}$
through the relation
\begin{align}
\Mclabdnjm{d}{njm} &= \sum_{m'} \Mccavdnjm{d}{njm'} D^{(j)}_{m'm}(-\ga,-\be,-\al) 
\notag \\
&= \sum_{m'} \Mccavdnjm{d}{njm'} e^{im'\ga} e^{im\al} d^{(j)}_{m'm}(-\be) \ ,
\label{rotations}
\end{align}
where $\al$, $\be$, and $\ga$
are the Euler angles
relating the cavity and laboratory frames.
The cavity frame is found by starting
with the two frames aligned,
then rotating the cavity frame
by $\al$ about the $z$ axis
followed by a rotation of
$\be$ about the new $y$ axis
and a rotation of $\ga$
about the new $z$ axis.

Most experiments involve cavities
that lie in the  horizontal plane.
This is achieved by taking
$\be = 90^\circ$.
Then $\al$ gives the angle
between the cavity axis and
the laboratory $x$ axis
and $\ga$ gives the angle
of the polarization out of the
horizontal plane.
The result is then
\begin{align}
\Mclabdnjm{d}{njm} &= 
\sum_{m'} \Mccavdnjm{d}{njm'}
e^{im'\ga} e^{im\al} d^{(j)}_{m'm}(-\tfrac\pi2)
\notag \\
&= \Mzero{d}{nj} e^{im\al} d^{(j)}_{0m}(-\tfrac\pi2) 
\notag \\&\quad
+ \Mtwo{d}{nj} e^{im\al} e^{i2\ga} d^{(j)}_{2m}(-\tfrac\pi2)
\notag \\&\quad
+ \Mtwo{d}{nj} e^{im\al} e^{-i2\ga} d^{(j)}_{(-2)m}(-\tfrac\pi2) \ .
\end{align}
Again, only even values of $j$
contribute in Fabry-P\'erot cavities.

The above expression can now be used
to find the modulation amplitudes
in the beat frequency.
Consider, for example,
the two recent experiments in Refs.\ \cite{schiller,peters}.
Both experiments involve orthogonal
Fabry-P\'erot cavities.
The cavities are empty,
so there will be no polarization-dependent
contribution from $\Mtwo{d}{nj}$.
Taking one cavity along the laboratory $x$ axis ($\al=0$)
and the other along the $y$ axis ($\al=\pi/2$),
variations in the beat frequency are controlled by
\beq
\De\Mclabdnjm{d}{njm} = \Mzero{d}{nj} d^{(j)}_{0m}(-\tfrac\pi2)
\big( 1 - i^m \big) \ .
\eeq
Notice that this vanishes
when $m$ is a multiple of 4.
This is because a $90^\circ$ rotation
about the vertical axis corresponds
to interchanging the two cavities.
Therefore, the beat frequency must change sign
under a $90^\circ$ rotation
and is invariant under a $180^\circ$ rotation.
This implies that only harmonics
with $m$ equal to an even value
not divisible by four can appear.

\begin{table*}
  \renewcommand{\tabcolsep}{5pt}
  \renewcommand{\arraystretch}{2.0}
  \begin{tabular}{c|cc|c|c|c|c}
    dimension & $m$ & $m'$ & $C^C_{mm'}$ & $C^S_{mm'}$ & $S^C_{mm'}$ & $S^S_{mm'}$ \\
    \hline\hline
    $d=4$ 
    & 2 & 0 & 
    $-0.36\, \cIdjm{4}{20}$ & 0 & 0 & 0 \\
    & 2 & 1 &
    $-0.75\, \re\Big[\cIdjm{4}{21}\Big]$ & $0.75\, \im\Big[\cIdjm{4}{21}\Big]$ &
    $0.95\, \im\Big[\cIdjm{4}{21}\Big]$ & $0.95\, \re\Big[\cIdjm{4}{21}\Big]$ \\
    & 2 & 2 &
    $-1.3\, \re\Big[\cIdjm{4}{22}\Big]$ & $1.3\, \im\Big[\cIdjm{4}{22}\Big]$ &
    $1.2\, \im\Big[\cIdjm{4}{22}\Big]$ & $1.2\, \re\Big[\cIdjm{4}{22}\Big]$ \\[4pt]
    \hline
    $d=6$ 
    & 2 & 0 & 
    $3.9\, \cfdnjm{6}{220}$ & 0 & 0 & 0 \\
    & 2 & 1 &
    $8.2\, \re\Big[\cfdnjm{6}{221}\Big]$ & $-8.2\, \im\Big[\cfdnjm{6}{221}\Big]$ &
    $-10\, \im\Big[\cfdnjm{6}{221}\Big]$ & $-10\, \re\Big[\cfdnjm{6}{221}\Big]$ \\
    & 2 & 2 &
    $14\, \re\Big[\cfdnjm{6}{222}\Big]$ & $-14\, \im\Big[\cfdnjm{6}{222}\Big]$ &
    $-13\, \im\Big[\cfdnjm{6}{222}\Big]$ & $-13\, \re\Big[\cfdnjm{6}{222}\Big]$ \\[4pt]
    \hline
    $d=8$
    & 2 & 0 & 
    $11\, \cfdnjm{8}{420}$ & 0 & 0 & 0 \\[-4pt]
    & & & 
    $-20\, \cfdnjm{8}{440}$ & & & \\
    & 2 & 1 &
    $22\, \re\Big[\cfdnjm{8}{421}\Big]$ & $-22\, \im\Big[\cfdnjm{8}{421}\Big]$ &
    $-29\, \im\Big[\cfdnjm{8}{421}\Big]$ & $-29\, \re\Big[\cfdnjm{8}{421}\Big]$ \\
    & & & 
    $-4.8\, \re\Big[\cfdnjm{8}{441}\Big]$ & $+4.8\, \im\Big[\cfdnjm{8}{441}\Big]$ &
    $+29\, \im\Big[\cfdnjm{8}{441}\Big]$ & $+29\, \re\Big[\cfdnjm{8}{441}\Big]$ \\
    & 2 & 2 &
    $38\, \re\Big[\cfdnjm{8}{422}\Big]$ & $-38\, \im\Big[\cfdnjm{8}{422}\Big]$ &
    $-37\, \im\Big[\cfdnjm{8}{422}\Big]$ & $-37\, \re\Big[\cfdnjm{8}{422}\Big]$ \\
    & & & 
    $+0.53\, \re\Big[\cfdnjm{8}{442}\Big]$ & $-0.53\, \im\Big[\cfdnjm{8}{442}\Big]$ &
    $-10\, \im\Big[\cfdnjm{8}{442}\Big]$ & $-10\, \re\Big[\cfdnjm{8}{442}\Big]$ \\
    & 2 & 3 &
    $23\, \re\Big[\cfdnjm{8}{443}\Big]$ & $-23\, \im\Big[\cfdnjm{8}{443}\Big]$ &
    $-20\, \im\Big[\cfdnjm{8}{443}\Big]$ & $-20\, \re\Big[\cfdnjm{8}{443}\Big]$ \\
    & 2 & 4 &
    $-16\, \re\Big[\cfdnjm{8}{444}\Big]$ & $16\, \im\Big[\cfdnjm{8}{444}\Big]$ &
    $16\, \im\Big[\cfdnjm{8}{444}\Big]$ & $16\, \re\Big[\cfdnjm{8}{444}\Big]$
  \end{tabular}
  \caption{\label{table_fabry_perot}
    Nonzero modulation amplitudes
    for the Fabry-P\'erot cavities
    in Refs.\ \cite{schiller,peters}.
    Camouflage coefficients
    up to dimension $d=8$ are included.
    The numbers $m$ and $m'$ give the harmonics
    of the turntable rotation frequency
    and sidereal frequency, respectively.
    The dimension-6 amplitudes are
    in units of $10^{-18}$ GeV$^2$.
    The dimension-8 amplitudes are
    in units of $10^{-36}$ GeV$^4$.}
\end{table*}

Both the experiments under consideration
use Nd:YAG lasers as the radiation source
($\om = 1.17\ \text{eV}$) and
lie at similar colatitudes, $\ch\simeq 38^\circ$.
Consequently, both experiments are
sensitive to nearly identical 
combinations of coefficients for Lorentz violation.
These combinations are the modulation amplitudes
from Eq.\ \rf{CSamps}.
They are given up to $d=8$ in Table \ref{table_fabry_perot}.
The table shows that we only expect
variations at twice the turntable rate.

While symmetry considerations 
prohibit turntable harmonics
for odd values of $m$
and multiples of four,
higher-order variations with $m\geq 6$
can arise in general.
Variations at six times the
turntable frequency do arise from operators
of dimension $d=10$ and higher, for example.
However, the absence of odd harmonics
and those at multiples of four times the turntable rate
is a generic feature of any experiment
based on two identical crossed cavities lying
in the horizontal plane.
Choosing a relative angle other than $90^\circ$
or orienting one cavity out of the horizontal plane
may provide additional sensitivity to Lorentz violation.

Comparing Table \ref{table_fabry_perot}
to the microwave experiment in
Ref.\ \cite{parker},
we confirm the enhanced
potential sensitivities arising
from the $\om^{d-4}$ dependence
in the frequency shift.
The table shows that the camouflage coefficients
enter with a factor on the order of
$10^{-18}\ \text{GeV}^2$ for $d=6$
and 
$10^{-36}\ \text{GeV}^4$ for $d=8$.
Assuming dimensionless sensitivities to variations
in the beat frequency on the order of $10^{-17}$,
we expect measurements of
camouflage coefficients at the level of 
$10 \text{ GeV}^{-2}$ for $d=6$
and 
$10^{19} \text{ GeV}^{-4}$ for $d=8$.
These sensitivities are many orders of magnitude
beyond the current microwave bounds.

\section{ring resonator} \label{sec_ring}

\begin{figure}
\vskip5pt
\begin{centering}
\includegraphics{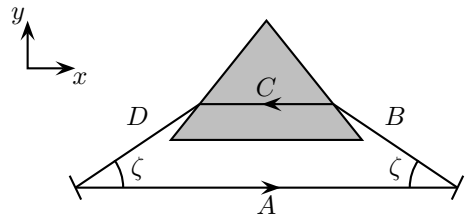}
\end{centering}
\caption{\label{ring_fig}
  Ring resonator composed of two mirrors and a prism
  with index of refraction $N$.
  The arms $A$, $B$, $C$, and $D$ are of length
  $L_A$, $L_B$, $L_C$, and $L_D$, respectively.
  Refraction at the prism is at Brewster's angle.
  Polarization is linear in the plane of the ring.
  Arm $A$ is oriented along the laboratory-frame
  $x$ axis.
}
\end{figure}

\begin{table*}
  \renewcommand{\tabcolsep}{5pt}
  \renewcommand{\arraystretch}{2.0}
  \begin{tabular}{c|cc|c|c|c|c}
    dimension & $m$ & $m'$ & $C^C_{mm'}$ & $C^S_{mm'}$ & $S^C_{mm'}$ & $S^S_{mm'}$ \\
    \hline\hline
    $d=6$ 
    & 1 & 0 & 
    $0.22\, \cfdnjm{6}{110}$ & 0 & 0 & 0 \\
    & 1 & 1 &
    $-0.20\, \re\Big[\cfdnjm{6}{111}\Big]$ & $0.20\, \im\Big[\cfdnjm{6}{111}\Big]$ &
    $-0.37\, \im\Big[\cfdnjm{6}{111}\Big]$ & $-0.37\, \re\Big[\cfdnjm{6}{111}\Big]$ \\[4pt]
    \hline
    $d=8$
    & 1 & 0 & 
    $-0.61\, \cfdnjm{8}{110}$ & 0 & 0 & 0 \\
    & & & 
    $+6.5\, \cfdnjm{8}{310}$ & & & \\
    & & & 
    $-1.5\, \cfdnjm{8}{330}$ & & & \\
    & 1 & 1 &
    $0.54\, \re\Big[\cfdnjm{8}{111}\Big]$ & $-0.54\, \im\Big[\cfdnjm{8}{111}\Big]$ &
    $1.0\, \im\Big[\cfdnjm{8}{111}\Big]$ & $1.0\, \re\Big[\cfdnjm{8}{111}\Big]$ \\
    & & & 
    $-5.8\, \re\Big[\cfdnjm{8}{311}\Big]$ & $+5.8\, \im\Big[\cfdnjm{8}{311}\Big]$ &
    $-11\, \im\Big[\cfdnjm{8}{311}\Big]$ & $-11\, \re\Big[\cfdnjm{8}{311}\Big]$ \\
    & & & 
    $-9.2\, \re\Big[\cfdnjm{8}{331}\Big]$ & $+9.2\, \im\Big[\cfdnjm{8}{331}\Big]$ &
    $+1.0\, \im\Big[\cfdnjm{8}{331}\Big]$ & $+1.0\, \re\Big[\cfdnjm{8}{331}\Big]$ \\
    & 1 & 2 &
    $-1.1\, \re\Big[\cfdnjm{8}{332}\Big]$ & $1.1\, \im\Big[\cfdnjm{8}{332}\Big]$ &
    $7.2\, \im\Big[\cfdnjm{8}{332}\Big]$ & $7.2\, \re\Big[\cfdnjm{8}{332}\Big]$ \\
    & 1 & 3 &
    $3.8\, \re\Big[\cfdnjm{8}{333}\Big]$ & $-3.8\, \im\Big[\cfdnjm{8}{333}\Big]$ &
    $7.1\, \im\Big[\cfdnjm{8}{333}\Big]$ & $7.1\, \re\Big[\cfdnjm{8}{333}\Big]$ \\
    & 3 & 0 &
    $2.5\, \re\Big[\cfdnjm{8}{330}\Big]$ & 0 & 0 & 0 \\
    & 3 & 1 &
    $-2.7\, \re\Big[\cfdnjm{8}{331}\Big]$ & $2.7\, \im\Big[\cfdnjm{8}{331}\Big]$ &
    $-5.0\, \im\Big[\cfdnjm{8}{331}\Big]$ & $-5.0\, \re\Big[\cfdnjm{8}{331}\Big]$ \\
    & 3 & 2 &
    $4.8\, \re\Big[\cfdnjm{8}{332}\Big]$ & $-4.8\, \im\Big[\cfdnjm{8}{332}\Big]$ &
    $4.0\, \im\Big[\cfdnjm{8}{332}\Big]$ & $4.0\, \re\Big[\cfdnjm{8}{332}\Big]$ \\
    & 3 & 3 &
    $-3.2\, \re\Big[\cfdnjm{8}{333}\Big]$ & $3.2\, \im\Big[\cfdnjm{8}{333}\Big]$ &
    $-3.3\, \im\Big[\cfdnjm{8}{333}\Big]$ & $-3.3\, \re\Big[\cfdnjm{8}{333}\Big]$
  \end{tabular}
  \caption{\label{table_ring}
    Nonzero modulation amplitudes
    for the ring resonator in Ref.\ \cite{ring}.
    Camouflage coefficients
    up to dimension $d=8$ are included.
    The numbers $m$ and $m'$ give the harmonics
    of the turntable rotation frequency
    and sidereal frequency, respectively.
    The dimension-6 amplitudes are
    in units of $10^{-18}$ GeV$^2$.
    The dimension-8 amplitudes are
    in units of $10^{-36}$ GeV$^4$.}
\end{table*}

Ring resonators provide another
example of an optical experiment
sensitive to Lorentz violation.
One advantage of a ring
resonator is that it is not
symmetric under parity.
As a result, it can provide
unsuppressed sensitivity to parity-odd
coefficients for Lorentz violation.
Here, we will consider the resonator
of Ref.\ \cite{ring} specifically,
but other configurations are possible.
The basic setup is shown in Fig.\ \ref{ring_fig}.
Plane waves polarized in the plane
of the oscillator propagate around
the resonator.
Resonant frequencies
for the clockwise and counterclockwise
modes are then compared
and analyzed for signatures of Lorentz violation.

Much of the analysis of the ring resonator
mirrors that of the Fabry-P\'erot cavity.
We begin by defining a beam frame where
the wave propagates along the $z$ axis
and is polarized in the $y$ direction.
The beam is then given by the nonzero component
$E_y(\vec x) = \calE\rh(\vec x)\exp ikz$,
where $\rh$ is the beam profile function.
The smooth field is taken by
extending the plane wave to infinity,
$\sE_y(\vec x) = \calE\exp ikz$.
Following the same steps as before,
we find that the $\Mc$ matrix elements
associated with radiation propagating
in one direction in a given arm
of the resonator is
\beq
\Mcarm = \Mzero{d}{nj} \de_{m,0} + \Mtwo{d}{nj} \de_{|m|,2} \ ,
\label{Marm}
\eeq
where, in this case,
$j$ takes on any value satisfying $j\geq |m|$.
The symmetries that prevented odd $j$ values
in the Fabry-P\'erot case no longer apply,
so all values of $j$ can contribute, in principle.
The $\Mzero{d}{nj}$ and $\Mtwo{d}{nj}$
factors are given in Eqs.\ \rf{Mzero} and \rf{Mtwo}.

The above calculation gives the contribution
from a single arm.
The total $\Mc$ matrix for a given mode
is the energy-weighted average of all the arms.
To find the average,
we must first rotate the single arm result
$\Mcarm$ to get the proper orientation in
the laboratory frame.

Consider the counterclockwise mode propagating
around the ring in Fig.\ \ref{ring_fig}.
For arms $A$, $B$, and $D$ we take $N=1$, 
which gives zero $\Mtwo{d}{nj}$.
The $\Mc$ matrix for arm $A$
is then obtained by a $\be = 90^\circ$ rotation
about the $y$ axis,
giving a beam propagating in
the laboratory-frame $x$ direction.
For $B$, the $\be=90^\circ$ rotation
is preceded by a $\al=180^\circ-\ze$
rotation about the $z$ axis.
Similarly,
the matrix for $C$ is given by
rotation angles $\al=180^\circ$ and $\be=90^\circ$,
and arm $D$ is found using $\al= 180^\circ + \ze$ and $\be=90^\circ$.
The resulting laboratory-frame
matrix elements can be written as
\begin{align}
\McarmA &= \MzeroA{A}{d}{nj} d^{(j)}_{0m}(-\tfrac\pi2) \ ,\db\\
\McarmB &= \MzeroA{B}{d}{nj}(-1)^m e^{-im\ze}
d^{(j)}_{0m}(-\tfrac\pi2) \ , \db\\
\McarmC &= \MzeroA{C}{d}{nj} (-1)^m d^{(j)}_{0m}(-\tfrac\pi2)
\notag\\ &\quad
%+ \MtwoA{C}{d}{nj} (-1)^m  d^{(j)}_{2m}(-\tfrac\pi2)
%\notag\\ &\quad
%+ \MtwoA{C}{d}{nj} (-1)^m  d^{(j)}_{(-2)m}(-\tfrac\pi2)
\hspace*{-30pt}
+ \MtwoA{C}{d}{nj} (-1)^m \Big( d^{(j)}_{2m}(-\tfrac\pi2)
+  d^{(j)}_{(-2)m}(-\tfrac\pi2) \Big)
\ ,\db\\
\McarmD &= \MzeroA{D}{d}{nj} (-1)^m e^{im\ze}
d^{(j)}_{0m}(-\tfrac\pi2) \ ,
\end{align}
where $\MzeroA{A}{d}{nj}=\MzeroA{B}{d}{nj}=\MzeroA{D}{d}{nj}$
is obtained by taking $N=1$ in Eq.\ \rf{Mzero},
while $\MzeroA{C}{d}{nj}, \MtwoA{C}{d}{nj}$ are found using
the index of refraction for the prism $N=N_\text{prism}$
in \rf{Mzero} and \rf{Mtwo}.

We next take the energy-weighted average of the four arms
to get the total laboratory-frame $\Mclab$ matrix.
Assuming perfect reflections at the mirrors
and Brewster's-angle refraction at the prism,
the power of the beam is the same in all four arms.
We may then write the power
$P = \la v = \la/N = \text{constant}$,
where $\la$ is the energy per length
of the beam.
The energy in a particular arm is
$\vev{U}_\text{arm} = \la L = P N L$.
This implies that the energy 
is proportional to the optical length $NL$. 
So, the energy fraction in a given arm
is the same as the fraction of the
total optical length attributed to that arm.
The laboratory-frame matrix
for the ring resonator is then given by
\begin{align}
\Mclab &= \frac{L_A}{L_\text{opt}} \McarmA
+\frac{L_B}{L_\text{opt}} \McarmB
\notag \\ &\quad
+\frac{NL_C}{L_\text{opt}} \McarmC
+\frac{L_D}{L_\text{opt}} \McarmD \ ,
\end{align}
where 
$L_\text{opt} = L_A + L_B + NL_C + L_D$
is the total optical path length.

Unlike the Fabry-P\'erot example,
where we compare identical modes
in two different cavities,
in this case we compare
two different modes in the same resonator.
The above calculation yields
the frequency shift for the
counterclockwise-propagating mode.
To get $\Mc$ for the clockwise mode,
we note that a reversal in propagation
in each arm is achieved through a 
$180^\circ$ rotation about the laboratory-frame $z$ axis.
This implies that the clockwise mode can be
found by multiplying the above result 
by $e^{im\pi} = (-1)^m$.
We then get
\beq
\De\Mclab = \big(1-(-1)^m\big) \Mclab \ .
\eeq
Notice that this vanishes for even values of $m$.
A $180^\circ$ rotation about the vertical
effectively interchanges the two modes,
changing the sign of the beat frequency.
As a result, no even harmonics 
of the turntable frequency can appear.
A parity transformation also interchanges the two modes,
changing the sign of the beat frequency.
Consequently, only parity-odd coefficients
for Lorentz violation should affect the beat frequency.
In the current context,
this corresponds to odd values of $j$.

The advantage of the ring resonator over
most other cavity experiments
is its sensitivity to parity-odd coefficients.
In order to demonstrate this sensitivity explicitly,
we consider the parameters
for the experiment in Ref.\ \cite{ring}.
The dimensions are
$L_A = 34.9$ mm, $L_B = L_D = 11.1$ mm, and $L_C = 14.1$ mm,
and the index of refraction of the prism is $N_\text{prism}=1.44$
\cite{tobar_pc}.
Again, the photon energy is approximately $\om = 1.17\ \text{eV}$.
The colatitude for this experiment is $\ch=122^\circ$.
The resulting modulation amplitudes
up to dimension $d=8$ are given in Table \ref{table_ring}.
Sensitivity to $j=1$ and $j=3$ coefficients is achieved, as expected.
Note, however,
that the mSME $d=4$ coefficients do not appear.

The absence of $d=4$ coefficients can be
understood by focusing on the
nonbirefringent parity-odd mSME terms.
There are a total of three coefficients in this limit,
which have previously been characterized
using an antisymmetric constant $3\times 3$ matrix
$\tilde\ka^{(4)}_{o+}$ \cite{km02}.
The nonzero components of 
$\tilde\ka^{(4)}_{o+}$ are linear combinations
of the three $\cIdjm{4}{1m}$ coefficients \cite{km09}.
The frequency shift due to these coefficients
can be written as
\beq
\frac{\de\nu}{\nu} = -\frac{1}{2\vev{U}} \, \ep^{jkl} \big(\tilde\ka^{(4)}_{o+}\big)^{kl}\,
\int d^3x\, \mu \vev{S^j} \ ,
\eeq
where
$\vev{\vec S} = \half \re\, \vec E^*\! \times \vec H$
is the time-averaged Poynting vector,
and $\mu$ is the permeability of the media.
For a constant permeability,
the frequency shift is proportional to 
the Poynting vector averaged over the volume
of the resonator,
which vanishes for a closed system.
Consequently, a lossless resonator cannot
provide sensitivity to these coefficients
unless media with different permeabilities are included.
We can show this explicitly in the ring resonator.
The Poynting vector in one arm can be written as
$\vev{\vec S} = P \vec L / V$,
where $P$ is the power,
$\vec L$ is the beam length vector, and
$V$ is the beam volume.
The frequency shift then becomes
\beq
\frac{\de\nu}{\nu} = \half \ep^{jkl} \big(\tilde\ka^{(4)}_{o+}\big)^{kl}\,
\frac{\sum_a \mu_a L_a^j}{L_\text{opt}} \ ,
\eeq
where we sum over the arms, $a=A,B,C,D$.
This is proportional to the vector
sum of the $\vec L_a$ vectors
in the event that the permeability $\mu$
is the same in all arms.
The sum vanishes for a closed path.

\begin{table}
\renewcommand{\arraystretch}{1.75}
\begin{tabular}{c||c|c||c|c|c}
coef.& F.P.& ring & P-even & P-odd & iso.\\
\hline
$\cIdjm{4}{jm}$   & 5  & 0  & 5  & 3  & 1 \\
%$\cfdnjm{4}{njm}$ & 0  & 0  & 0  & 0  & 1 \\
$\cfdnjm{6}{njm}$ & 5  & 3  & 5  & 3  & 2 \\
$\cfdnjm{8}{njm}$ & 13 & 12 & 19 & 13 & 3
\end{tabular}
\caption{\label{table_summary}
  Numbers of combinations of
  $d=4,6,8$ camouflage coefficients
  accessible to the existing
  experiments discussed in this work.
  The second column gives
  the number of parity-even 
  coefficients accessible to the
  Fabry-P\'erot experiments
  of Ref.\ \cite{schiller,peters}.
  The third column gives the number
  of parity-odd coefficients
  for the ring resonator of
  Ref.\ \cite{ring}.
  For comparison,
  the next three columns give
  the number of parity-even anisotropic,
  parity-odd anisotropic, and isotropic
  camouflage coefficients.} 
\end{table}

\section{discussion}\label{sec_discussion}

The results obtained in the two
examples given here are easily
generalized to other optical
experiments with linear polarization.
Assuming the resonator modes can
be decomposed into the superposition of plane waves,
we can use Eq.\ \rf{Marm} to get the
contribution from each wave.
We then use Wigner matrices, as in Eq.\ \rf{rotations},
to rotate the result for each component wave,
giving it the correct orientation 
in the laboratory.
Taking the energy-weighted average of the waves,
we arrive at the laboratory-frame
$\Mc$ matrix for the resonator.

As an example,
a Fabry-P\'erot cavity aligned with
the laboratory $x$ axis can be
treated as two plane waves at angles $\al = 0,\pi$.
A cavity aligned with the $y$ axis has $\al = \half\pi,\tfrac32\pi$.
For index of refraction $N=1$,
the difference is then
\begin{align}
\De\Mclab 
&= \half\big(1  + e^{im\pi} - e^{im\pi/2} - e^{im3\pi/2}\big)
\notag \\
&\qquad\times
\Mzero{d}{nj} d^{(j)}_{0m}(-\tfrac\pi2)\notag \\
%&= \half (1  + (-1)^m  - i^m - (-i)^m)
%\Mzero{d}{nj} d^{(j)}_{0m}(-\tfrac\pi2)\notag \\
&\hspace*{-20pt}
= \half (1  - i^m)(1+ (-1)^m)
\Mzero{d}{nj} d^{(j)}_{0m}(-\tfrac\pi2)\ ,
\end{align}
which vanishes unless $m$
is an even integer not divisible by four.
Using this and the property that 
$d^{(j)}_{0m}(-\tfrac\pi2)$ vanishes
unless $j$ and $m$ are
either both even or both odd,
we arrive at the same result obtained in
Sec.\ \ref{sec_fabry_perot}
for crossed Fabry-P\'erot cavities.

The above methods can be
applied to new optical experiments
utilizing different configurations.
These may provide sensitivity to
different combinations of coefficients
for Lorentz violation.
The results of this work demonstrate that,
while a large portion of the coefficient space
is accessible to the two types of experiment discussed here,
other experiments are needed to fully
constrain the coefficients up to $d=8$.
Table \ref{table_summary} summarizes the results.
A given Fabry-P\'erot experiment
based on orthogonal cavities
can measure twenty-three of
the twenty-nine anisotropic parity-even
camouflage coefficients.
A single ring resonator can test all but four of
the nineteen parity-odd coefficients.

For both classes of Lorentz test,
sensitivity to the missing anisotropic coefficients
could be achieved by performing experiments at
different latitudes
or by changing the experimental configuration.
For example, 
introducing matter with index
of refraction $N\neq 1$
in a Fabry-P\'erot experiment
or choosing different
cavity orientations will lead
to different sensitivities.
In principle,
all higher-order anisotropic coefficients
can be probed, without boost suppression,
using combinations of different
optical-cavity experiments.

Similar results are expected
for higher dimensions, $d=10,12,\ldots$,
which are easily included in an analysis.
One could also consider the larger
class of nonbirefringent coefficients $\cF$,
which include dispersive effects.
Only a few bounds from
astrophysical dispersion exist at present.
So, cavity experiments currently offer
an opportunity for an indirect
search for this unconventional feature
at interesting sensitivities.
Future analyses could also take advantage
of boost effects to probe
the isotropic coefficients.
These tests would see suppressions in sensitivity
of roughly $\be^2 \sim 10^{-8}$ in a parity-even experiment
and $\be \sim 10^{-4}$ in a parity-odd experiment.
Nonetheless, they may offer the best opportunity for
searches for these elusive forms of Lorentz violation.

\end{document}